\documentclass[twocolumn,notitlepage,nofootinbeib]{revtex4-2}
\usepackage{amsmath,amssymb}
\usepackage{graphicx}
\usepackage{newtxtext}
\usepackage{newtxmath}
\usepackage{verbatim}
\usepackage[x11names]{xcolor}
\usepackage[unicode,colorlinks,citecolor=Blue3,linkcolor=Blue3,bookmarks=true]{hyperref}
\usepackage[normalem]{ulem}
\usepackage{multirow}

\newcommand\bra[2][]{#1\langle {#2} #1|}
\newcommand\ket[2][]{#1|{#2} #1\rangle}
\newcommand{\braket}[2]{ \langle #1 | #2 \rangle}
\newcommand{\mean}[1]{\langle #1 \rangle}

\begin{document}

	\title{Preparation of Schr\"odinger cat quantum state using parametric down-conversion interaction}

	\author{V.\,L.\,Gorshenin}
	\email{valentine.gorshenin@yandex.ru}
	\affiliation{Russian Quantum Center, Skolkovo 121205, Russia}
	\affiliation{Moscow Institute of Physics and Technology, 141700 Dolgoprudny, Russia}

	\begin{abstract}

		The Schr\"odinger cat (SC) states are important in quantum optics because of their non-Gaussian properties. We propose a novel method of conditional generation of bright (multi-photon) SC states that uses degenerate parametric down-conversion and heralding measurement of the photon number in the pump mode. We show that this method, in principle, could be implemented using the modern high-\(Q\) optical microresonators.

	\end{abstract}

	\maketitle

  \section{Introduction}

Non-Gaussian quantum states, that is quantum states of continuous-variable (CV) systems \cite{Braunstein_RMP_77_513_2005}, described by negative-valued Wigner  quasi-probability functions \cite{Wigner_PR_40_749_1932,R_L_Hudson_Reports_Mat_Phys_6_2_1974, Schleich2001},
can be considered as ``more quantum'' that other ones. They are of special interest to fundamental tests of applicability of quantum physics to macroscopic objects \cite{W_Marshall_PRL_91_14_2003, O_Romero-Isart_NewJofPhys_12_3_2010, F_Khalili_PRL_105_7_2010}, to emerging quantum information technologies	\cite{T_C_Ralph_PRA_68_4_2003, T_C_Ralph_PhysRevLett_95_10_2005, G_Y_Xiang_NatPhotonics_4_5_2010, B_P_Lanyon_NatPhys_5_2_2009, A_P_Lund_PRL_100_3_030503_2008, A_I_Lvovsky_Arxiv_2006_16985_2020, M_Walschaers_PRX_Quant_2_3_2021}, and to high-precision interferometric measurements \cite{Shukla_OptQEl_55_460_2023, Shukla_PhOpen_18_100200_2024, Singh_PhOpen_18_100198_2024, V_L_Gorhsenin_21_6_2024, V_L_Gorshenin_Arxiv_2405_07049}.

  One of the most well known type of the non-Gaussian states is the Schr\"odinger cat (SC) one.  The ``mainstream'' method of preparation of these states is based in photon subtraction and conditional measurement, see Refs.\,\cite{M_Dakna_PRA_55_4_1997, J_S_Neergaard_PRL_97_8_2006, A_Ourjoumtsev_Science_312_5770_2006, Ourjoumtsev_A_Nat_448_7115_2007, K_Wakui_OptExpress_15_6_2007, K_Huang_PRL_115_2_2015, Sychev_NaturePhotonics_11_379_2017} and the review \cite{A_Lvovsky_Arxiv_2006_16985}. The values of SC state amplitudes up to $\alpha\sim2$ were demonstrated using this method. Iterative growth of the state of SC was also proposed and demonstrated \cite{Etesse_J_PRL_114_19_2015, Sychev_NaturePhotonics_11_379_2017}. The recent estimates \cite{D_A_Kuts_PhyicaScripta_97_11_2022, M_S_Podoshvedov_SR_13_3965_2023} show that it allows to prepare SC states with the amplitudes \(\alpha \gtrsim 4\) and fidelity exceeding 99\%. We would like to mention also the experimental works aimed on increasing the probability of preparation of the SC state \cite{Wang_M_Laser_n_Photonic_Reviews_16_12_2022} and on the remote preparation of the SC state using two-mode squeezing \cite{Dongmei_H_Laser_n_Photonic_Reviews_17_9_2023}. The influence of optical losses on the SC state was explored experimentally in Ref. \cite{Zhang_PhotRes_9_927_2021}.

  It was shown in the works \cite{S_P_Nikitin_IOP_1991, R_Tanas_IOP_1991} that much more bright quantum states having the characteristic SC-like two maxima shape can be generated using the second order optical nonlinearity described by the Hamiltonian
	\begin{equation} \label{SHG-hamiltonian}
		\hat{H}_2 = \hbar\gamma
      \left(\hat{a}^2 \hat{b}^\dagger + \hat{a}^{\dagger 2}\hat{b}\right) ,
	\end{equation}
	where \(\gamma\) is the parametric interaction coefficient and \(\hat{a}\), \(\hat{b}\) are the annihilation operator of two optical modes  with the frequencies,  equal to, respectively, $\omega_s$ (the signal mode) and $\omega_p=2\omega_s$ (the pump mode).

  Depending on the initial conditions, this Hamiltonian could describes nonlinear processes of the second harmonic generation or the degenerate parametric down-conversion. It is the second one that is responsible for the SC-like state generation in the signal mode.

	In particular, it was shown in Refs.\,\cite{P_D_Drummond_JPhysA_13_7_1980, P_D_Drummond_Optica_Acta_27_3_1980} that the degenerate parametric down-conversion can  effectively transform the losses in the pump mode to two-photon losses in the signal one. In this case, a pair of photons from the signal mode is transformed into one photon of the pump mode that then dissipates. It was shown that in the absence of ordinary single-photon losses in the signal mode, the SC state is a steady-state of this system \cite{L_Gilles_PRA_48_2_1993, L_Gilles_PRA_49_4_1994, E_E_Hach_III_PRA_49_1_1994}. However, the ordinary linear losses in the signal mode make the SC state only transient.

  Unfortunately, this method requires a very strong parametric interaction and small linear losses in the signal mode \cite{M_Wolinsky_PRL_60_18_1988}. By now, these conditions were successfully fulfilled only in microwave cryogenic systems with Josephson junctions \cite{Z_Leghtas_Science_347_6224_2015, R_Y_Teh_PRA_101_4_2020, Han_K_OptExpress_29_9_2021}. Cold-trapped ions \cite{Ding_PRL_119_15_2017} and semiconductor microcavities \cite{Arka_M_PRB_87_23_2018, Shen_H_PRA_90_2_2014, Ye_C_PRA_84_5_2011} can also be used as possible platforms for various applications of the spontaneous parametric-down conversion.

  The joint unitary evolution of the two modes was analyzed in the works \cite{S_P_Nikitin_IOP_1991, R_Tanas_IOP_1991, Singh_R_ArXiv_2405_14526} numerically, showing that quantum states that resemble SC ones (but not equal to them) can be generated in this process; see more details in Sec.\,\ref{sec:naive-approach}.

  In this work, we propose a new method of preparation of bright (multi-photon) SC states based on the degenerate parametric down-conversion process and the heralding measurement of the photon number in the pump mode.

  The paper is organized as follows. In Sec.\,\ref{sec:total-energy-repr} we review the convenient representation of the two-mode quantum states that was introduced in Refs.\,\cite{S_P_Nikitin_IOP_1991, R_Tanas_IOP_1991} and that we use in our numerical calculations here. In Sec.\,\ref{sec:naive-approach} we discuss the unconditional evolution of our system. In Sec.\,\ref{sec:her-meas} we introduce the conditional  preparation procedure and show that in principle, it allows to prepare the SC states with the fidelity approaching one. In Sec.\,\ref{sec:possible-implementation} we  summarize the results of this work and briefly discuss the possible implementation of the proposed procedure using the high-$Q$ optical microresonators.

	\section{Solution of of the Shr\"odinger equation}\label{sec:total-energy-repr}

	An analytical solution to the Shr\"odinger equation with the Hamiltonian \eqref{SHG-hamiltonian},
  \begin{equation}
    i \hbar \frac{\partial\ket{\Psi(t)}}{\partial t} = \hat{H}_2\ket{\Psi(t)} \,,
  \end{equation}
  is unknown. Therefore, we use the numerical solution in this paper. The straightforward approach in this case is to use the two-dimensional Fock basis
  \begin{equation}\label{nk}
    \{\ket{n}_s\otimes\ket{n}_p\} \,,
  \end{equation}
  where $\ket{n}_s$, $\ket{n}_p$ are the energy eigenstates of the signal and pump modes.
  However, numerical calculation using this approach, especially in the most interesting case of the multi-photon states, requires very significant computational resources.

  At the same time, note that the sum energy of the two modes
  \begin{equation}
    \hat{H}_0 = \hbar\omega_s(\hat{a}^\dag\hat{a} + 2\hat{b}^\dag\hat{b})
  \end{equation}
  commutes with the Hamiltonian \eqref{SHG-hamiltonian}. This feature allows to represent $\hat{H}_2$ as follows:
  \begin{equation}
    \hat{H}_2 = \hbar\gamma\sum_{N=0}^\infty\sum_{k,k'=0}^{[N/2]}
      \ket{N-2k}_s \otimes \ket{k}_p H^{N}_{kk'}\,{}_s\!\bra{N-2k'}\otimes{}_p\!\bra{k'} \,,
  \end{equation}
  where [...] denotes the integer part of a number. The matrices $H^{N}_{kk'}$ have the following tri-diagonal form:
	\begin{equation} \label{tridialgonal-hamiltonian}
		\|H^{N}_{kk'}\| =
		\begin{bmatrix}
			0   	& c_0^N	& 0   	&\dots 	& 0 & 0\\
			c_0^N 	& 0		& c_1^N 	&\dots 	& 0 & 0\\
			0   	& c_1^N	& 0   	&\dots 	& 0 & 0\\
			\vdots 	&\vdots &\vdots &\ddots &
			c_{\left[\frac{N}{2}\right]-2}^N & 0\\
			0 		& 0 	& 0 	& c_{\left[\frac{N}{2}\right]-2}^N
			& 0 & c_{\left[\frac{N}{2}\right]-1}^N\\
			0 		& 0 	& 0 	&0 		& c_{\left[\frac{N}{2}\right]-1}^N & 0\\
		\end{bmatrix} ,
	\end{equation}
	with
	\begin{equation}
		c^N_k = \sqrt{(k+1)(N-2k)(N-2k-1)} \,,
	\end{equation}
  see Refs.\,\cite{S_P_Nikitin_IOP_1991, R_Tanas_IOP_1991}.
  This ``semi-diagonalized'' representation significantly reduces the requirements for computational resources.

	Numerical methods for finding the eigenvectors of tridiagonal matrices are well developed. Let $\bigl\{\chi_k^{Nj}\bigr\}$, with $j=0\dots\left[\frac{N}{2}\right]$, be these  eigenvectors, and $\lambda^{Nj}$ -- the corresponding eigenvalues. Here the subscript $k$ enumerates the elements of the eigenvectors, and the pair of superscripts $N,j$ --- the eigenvectors and the corresponding eigenvalues. It is easy to see that in this case, the eigenstates of the Hamiltonian $\hat{H}_2$ are equal to
  \begin{equation}\label{chi_Nj}
    \ket{\chi^{Nj}}
    = \sum_{k=0}^{\left[\frac{N}{2}\right]}\chi_k^{Nj}\ket{N-2k}_s\otimes\ket{k}_p
  \end{equation}
  and the corresponding eigenvalues --- to $\hbar\gamma\lambda^{Nj}$.

  Let $\ket{\Psi}$ be the initial quantum state of our system. Using the basis $\{\ket{\chi^{Nj}}\}$, it can be presented as follows:
  \begin{equation}
    \ket{\Psi} = \sum_{N=0}^\infty\sum_{j=0}^{[N/2]}\Psi^{Nj}\ket{\chi^{Nj}} \,.
  \end{equation}
  Its time evolution is described by the following equation:
	\begin{equation}
		\ket{\Psi(t)} = e^{-i\hat{H}_2t/\hbar}\ket{\Psi}
    = \sum_{N=0}^{N=\infty}\sum_{j=0}^{[N/2]}e^{-i\tau\lambda^{Nj}}
        \Psi^{Nj}\ket{\chi^{Nj}}  \,,
	\end{equation}
  where
  \begin{equation}\label{eq:dimensionless-time-def}
    \tau = \gamma t
  \end{equation}
  is the normalized time.


	\section{Unconditional evolution}\label{sec:naive-approach}

\begin{figure}
		\centering
		\includegraphics[width=0.9\linewidth]{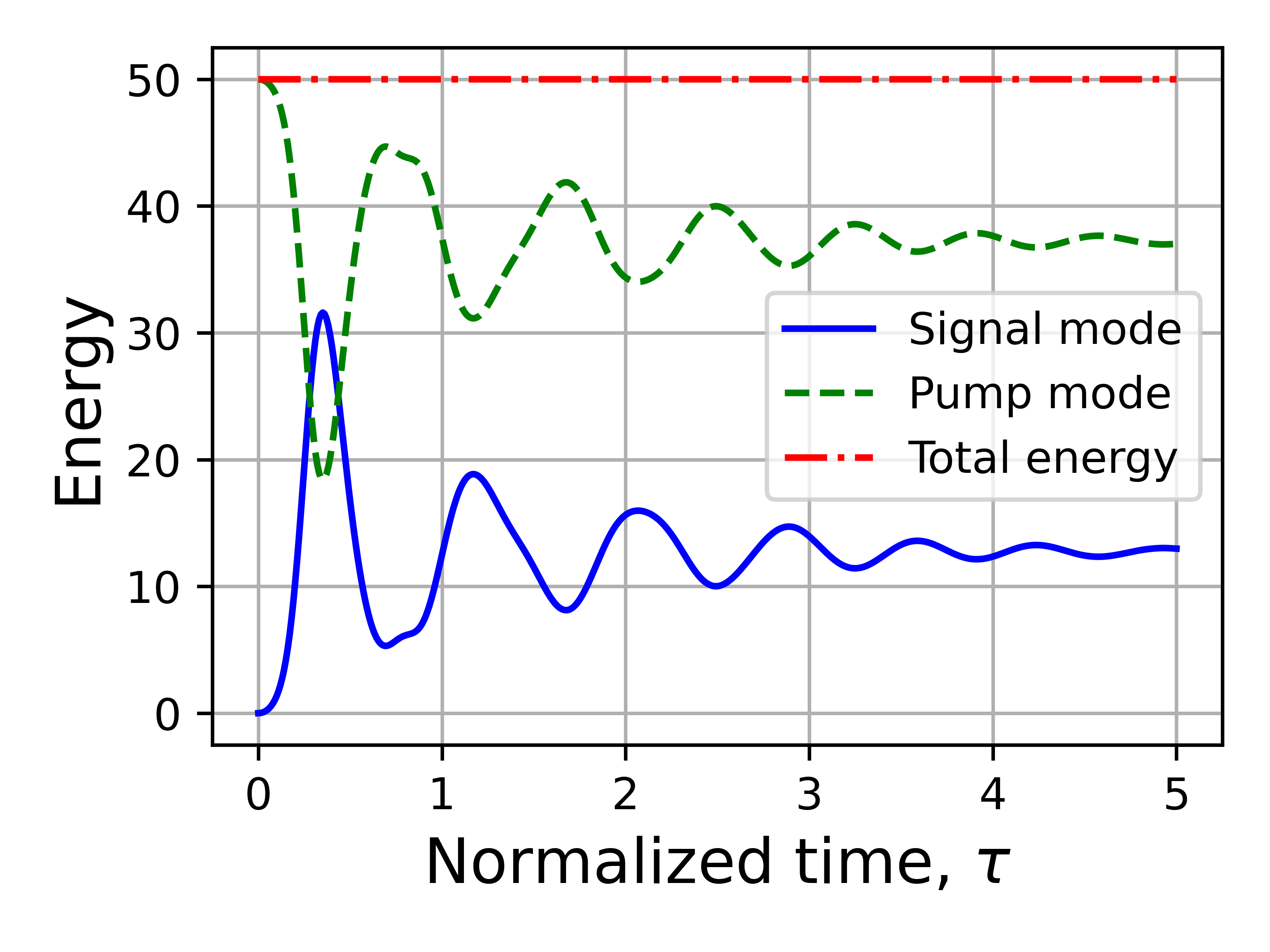}
		\caption{Blue solid: normalized mean energy of the signal mode $\mean{\hat{n}_s}$; green dashed line: normalized mean energy of the pump mode $2\mean{\hat{n}_p}$; red dash-dot line: the sum mean energy. The initial photon number in the pump mode $\mean{\hat{n}_p}=25$}
		\label{fig:ptrace-photon-num-dynamic}
	\end{figure}

	Suppose that initially the pump mode is prepared in the coherent quantum state \(\ket{\beta}_p\) and the signal mode --- in the vacuum state \(\ket{0}_s\). In Fig.\,\ref{fig:ptrace-photon-num-dynamic}, the corresponding mean values of the normalized energy of the signal mode $\mean{\hat{n}_s}$ and the pump one $2\mean{\hat{n}_p}$, calculated using the method considered in Sec.\,\ref{sec:total-energy-repr}, are plotted as the function of the normalized time $\tau$ for the initial photon number in the pump mode $\mean{\hat{n}_p}=25$. In addition, the sum mean energy $\mean{\hat{n}_s}+2\mean{\hat{n}_p}$ is also shown. It remains constant, since the total energy operator \(\hat{H}_0\) and the Hamiltonian \(\hat{H}_2\) commute.

  This plot clearly shows the interplay of the degenerate parametric down conversion process (which transfers the energy from the pump mode to the signal one) and the second harmonic generation one (which transfers the energy in the opposite direction). It is natural to expect that the maxima of $\mean{\hat{n}_s}$ correspond to the non-trivial non-Gaussian quantum states of the signal mode. This conclusion is supported also by Figs.\,1 and 4 of the paper \cite{S_P_Nikitin_IOP_1991}.

  In Ref.\,\cite{R_Tanas_IOP_1991}, evolution of the partial Husimi quasi-probability function of the signal mode was explored numerically. In Ref.\,\cite{Singh_R_ArXiv_2405_14526}, the partial Wigner functions of both the signal and the pump modes were calculated numerically for the small values of $\alpha\sim2$.  However, it has to be emphasized that in both these works, the resulting states were obtained by tracing out the second modes. This procedure gives inherently mixed states that are not so interesting for perspective applications of the non-Gaussian states.

\begin{figure}
		\centering
		\includegraphics[width=0.9\linewidth]{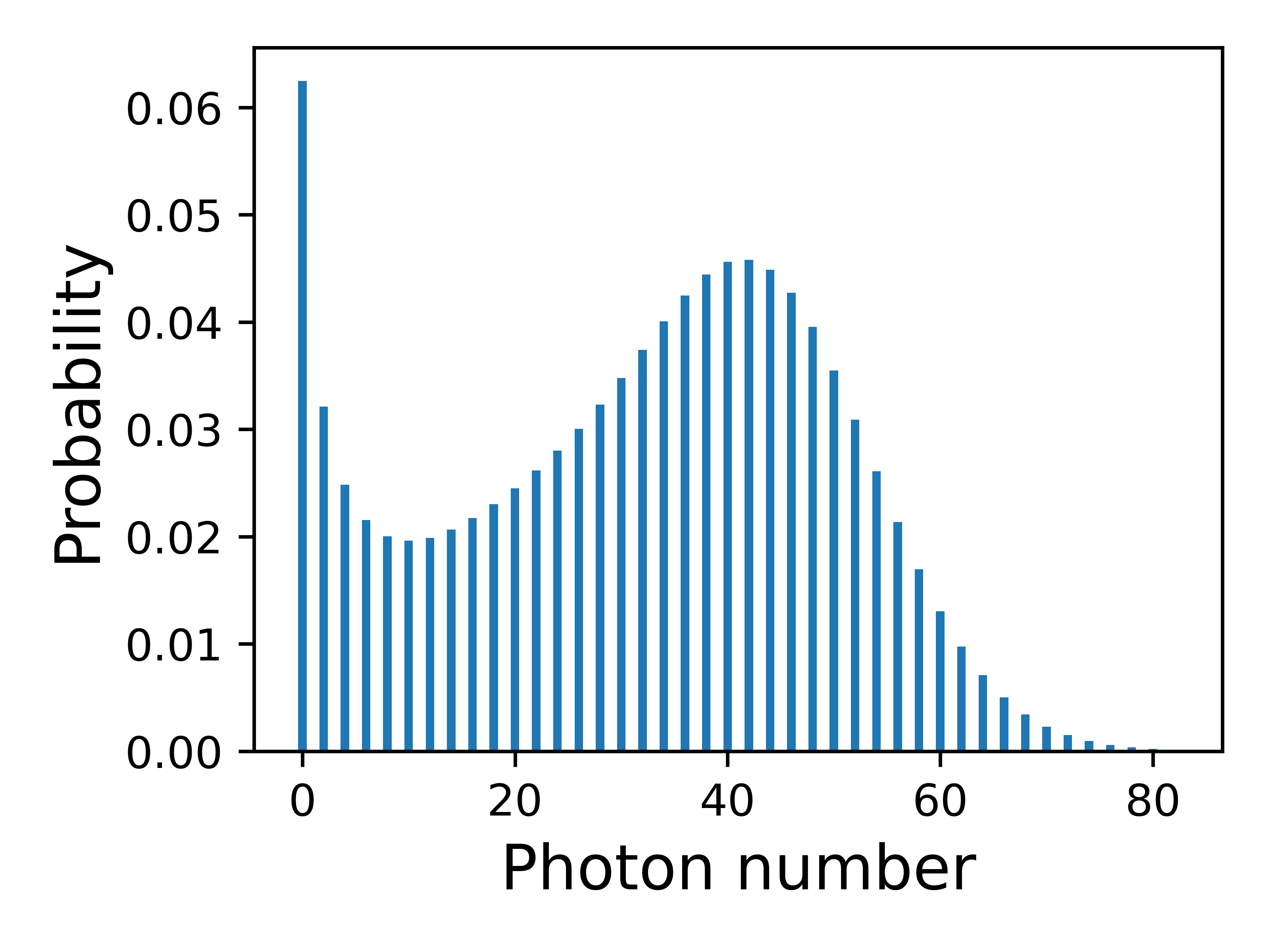}
		\caption{Probability distribution for the photon number in the signal mode at the optimal moment of time, showing the even statistics. The initial photon number in the pump mode $\mean{\hat{n}_p}=25$.}
		\label{fig:ptrace-bars-max-time}
	\end{figure}

  In Fig.\,\ref{fig:ptrace-bars-max-time}, the corresponding probability distribution for the photon number in the signal mode is shown. Similar to the ``even'' SC state
  \begin{equation}\label{SC}
   	\ket{\psi_{\rm cat}} = \frac{1}{\sqrt{K}}\Big(\ket{\alpha} + \ket{-\alpha}\Big) ,
  \end{equation}
  where $K$ is the normalization factor, the probabilities of the odd photon numbers are equal to zero, and the part of the distribution with the photon numbers above the mean value resembles the one of SC state. Unfortunately, this is not the case for the small values of $n_s$.

	\section{Conditional preparation of the Schr\"odinger cat state}\label{sec:her-meas}

\begin{figure}
		\centering
		\includegraphics[width=0.95\linewidth]{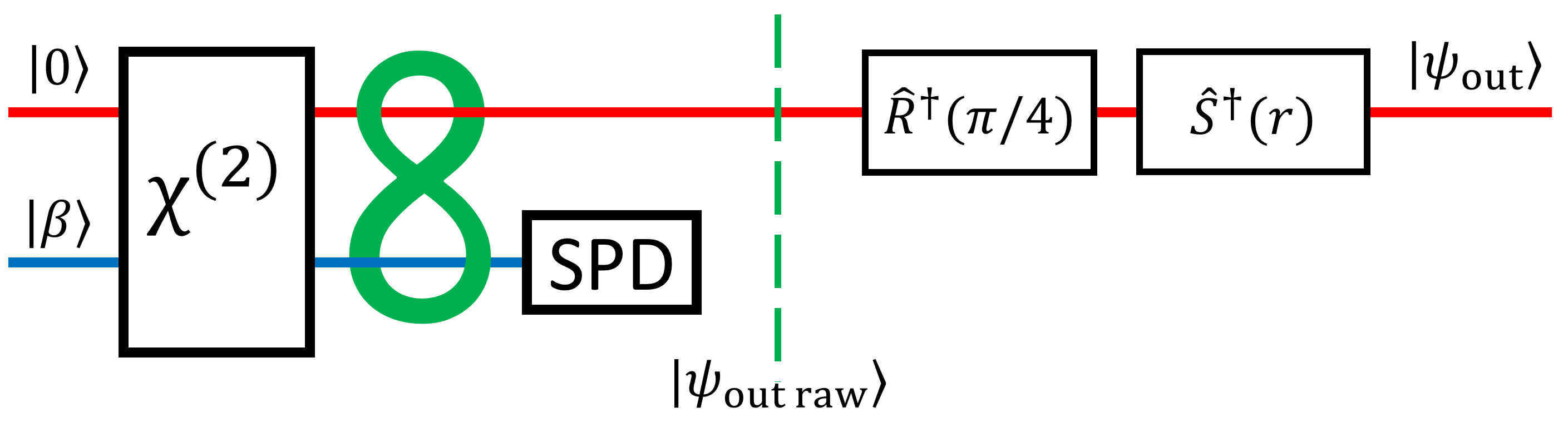}
		\caption{The scheme of the conditional preparation of the SC state. Red line -- signal mode, blue line -- pump mode. SPD -- single photon detector. The auxiliary rotation ($\hat{R}^\dag$) and anti-squeezing ($\hat{S}^\dag$) elements are discussed in the end of Sec.\,\ref{sec:her-meas}}.
		\label{fig:optic-scheme}
	\end{figure}

  Here we consider the protocol that, in principle, generate the pure quantum state approaching the SC one. Suppose that at some moment of the  normalized time $\tau_{\rm opt}$, the photon number \(\hat{n}_p\) in the pump mode is measured. Suppose also that we are interested only in the events where the value \(\hat{n}_p=0\) is obtained. The schematic diagram of this procedure is shown in Fig.\,\ref{fig:optic-scheme}.

	As a result, the joint state of our two-mode system reduces to the separable one
	\begin{equation}
		\ket{\Psi} = \ket{0}_p\otimes\ket{\psi_{\rm out}} \,,
	\end{equation}
	where $\ket{\psi_{\rm out}}$ is the resulting quantum state of the signal mode.

\begin{figure}
		\centering
		\includegraphics[width=0.9\linewidth]{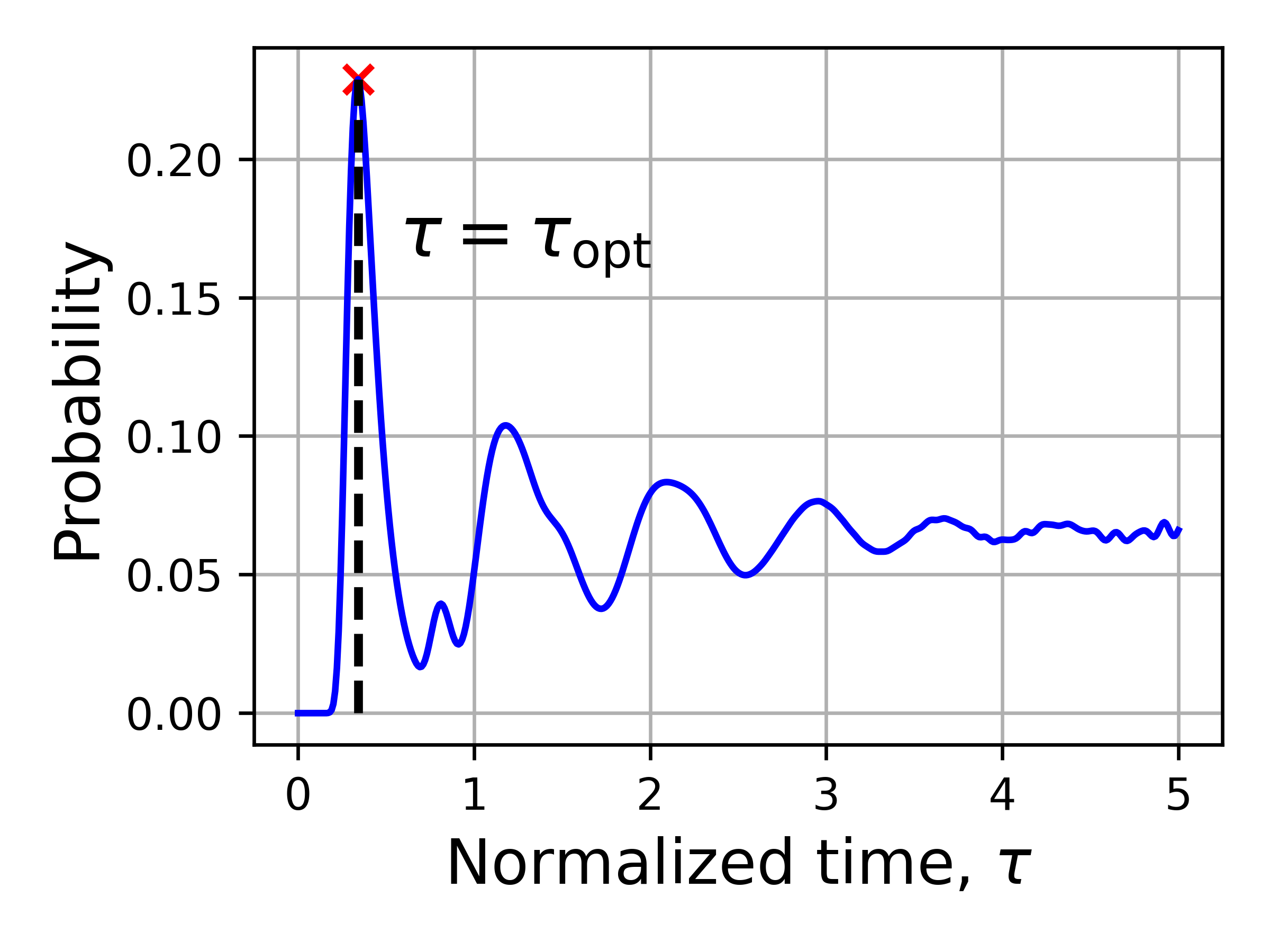}
		\caption{
			Dependence of the probability to measure zero photons in the pump mode on the
			interaction time. The initial photon number in the pump mode $\mean{\hat{n}_p}=25$.
		}
		\label{fig:prob-dynamics-her}
	\end{figure}

	We assume that the normalized time \(\tau_{\rm opt}\) corresponds to the maximum of the probability \(p_0\) of measuring zero photons in the pump mode. In Fig.\,\ref{fig:prob-dynamics-her}, this probability is plotted as a function of the normalized interaction time \(\tau\).

  The value of \(\tau_{\rm opt}\) decreases with the increase of the initial coherent state amplitude in the pump mode \(\beta\). This dependence can be approximated as follows:
	\begin{equation}\label{eq:t-max-by-beta}
		\tau_{\rm opt} = \gamma t_{\rm opt} \approx \frac{b_t}{(1+c_t \beta)^{d_t}}
	\end{equation}
	where \(b_t \approx 1.70\), \(c_t = 1.16\), and \(d_t \approx 0.84\). This equation, as well as the subsequent ones \eqref{eq:prob-approx},  \eqref{eq:sqeueeze-factor-approx}, are valid for the values of \(\beta\) within the range \([2, 100]\).

	Dependence of the success probability \(p_0\) on \(\beta\) can be approximated by the following function:
	\begin{equation}\label{eq:prob-approx}
		p_0 \approx \frac{b_p}{(1 + c_p  \beta)^{d_p}}
	\end{equation}
	where \(b_p \approx 2.56\), \(c_p \approx 1.95\), \(d_p = 1.02\).

\begin{figure}
		\centering
		\includegraphics[width=0.9\linewidth]{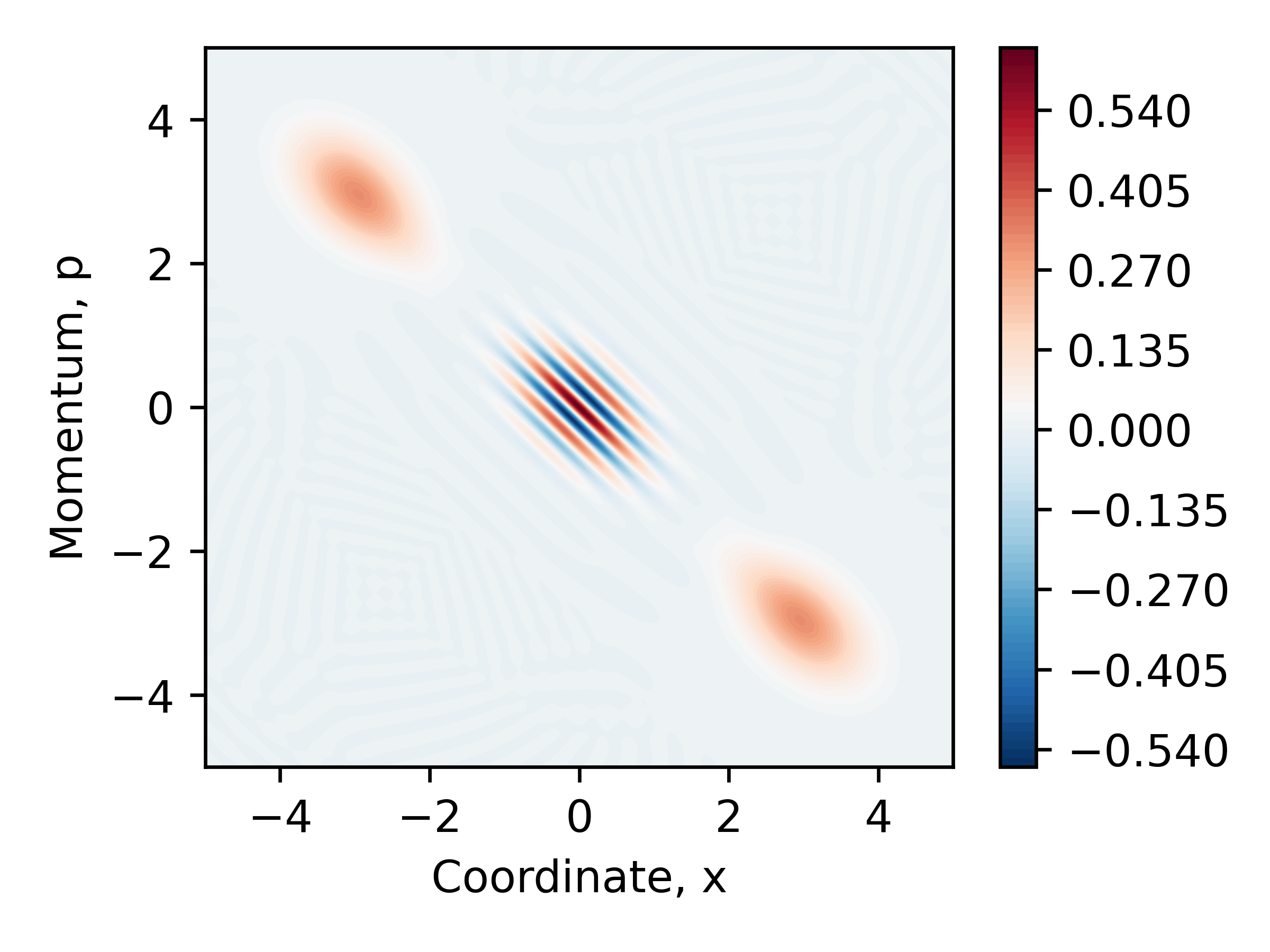}
		\caption{Wigner function of the output state \(\ket{\psi_{\rm out\,raw}}\) at the optimal time \(\tau_{\rm opt}\). \(\beta = 3\)}
		\label{fig:her-wigner-no-phase-shift}
	\end{figure}

	In Fig.\,\ref{fig:her-wigner-no-phase-shift}, the Wigner function of the resulting state $\ket{\psi_{\rm out\,raw}}$ is shown. It is easy to note that it strongly resembles the Wigner function of the squeezed and rotated by \(\pi/4\) SC state.

	As the quantitative measure of similarity of the obtained quantum state to the squeezed SC one, we consider the fidelity
	\begin{equation}\label{F}
		F = |\braket{\psi_{\rm out}}{\psi_{\rm cat}}|^2 \,,
	\end{equation}
	where $\ket{\psi_{\rm cat}}$ is the cat state \eqref{SC},
  \begin{equation}\label{psi_out}
    \ket{\psi_{\rm out}}
    =
    \hat{S}^\dag(\xi_{\rm prep})
    \hat{R}^\dag(\pi/4)
      \ket{\psi_{\rm out\,raw}} \,,
  \end{equation}
  \(\hat{R}\) is the unitary rotation operator,
  \begin{equation}
    \hat{S}(z) = \exp\left(\frac{1}{2}(z^* \hat{a}^2 - z \hat{a}^{\dagger 2})\right) \,,
  \end{equation}
  is the squeeze operator, and $\xi_{\rm prep}<0$ is the squeeze factor.

  The amplitude of prepared SC state \(\alpha_{\rm prep}\) is calculated in App.\,\ref{sec:app-sc-state-ampl-deriv}:
  \begin{equation}\label{eq:amplitude-of-prep-sc}
  	\alpha_{\rm prep} = e^{\xi_{\rm prep}} \sqrt{2 \beta^2 - \sinh^2(\xi_{\rm prep})} \,.
  \end{equation}
	(this equation is valid for the negative values of \(\xi_{\rm prep}\)).
	The optimal value of the squeeze parameter that provide the best fidelity can be approximated as follows:
	\begin{equation} \label{eq:sqeueeze-factor-approx}
		\xi_{\rm prep} \approx a_r + \frac{b_r}{(1 + c_r\beta)^{d_r}} \,,
	\end{equation}
	where \(a_r \approx -0.35\), \(b_r \approx 0.14\), \(c_r \approx 0.13\) and \(d_r = 2.40\). For considered here \(\beta > 0\) squeeze factor \(\xi_{\rm prep} < 0\).

	\begin{figure}
		\centering
		\includegraphics[width=0.9\linewidth]{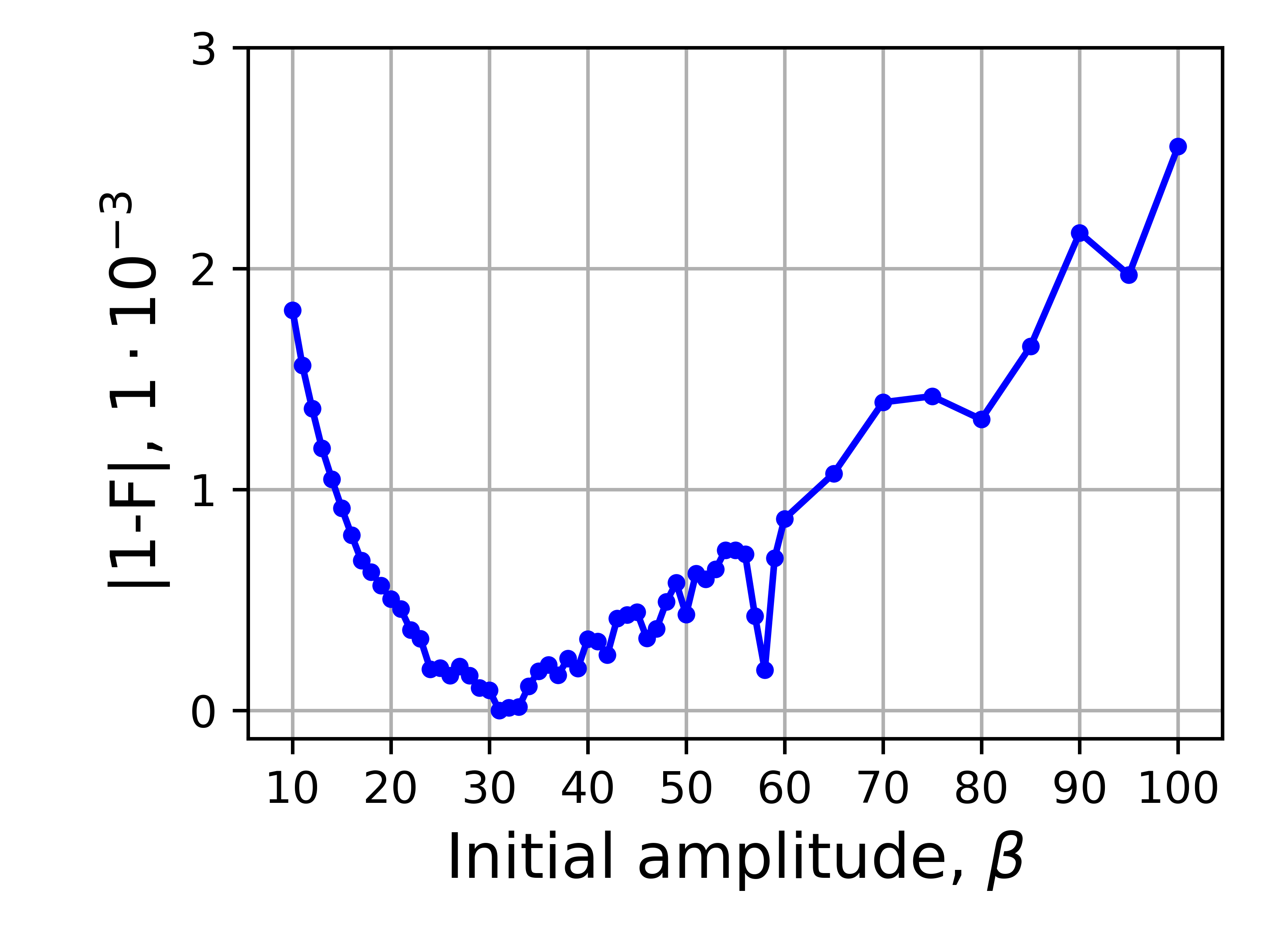}
		\caption{Deviation of the fidelity \eqref{F} from unity as a function of the amplitude in of the pump mode state \(\beta\).}
		\label{fig:her-overlap-plot-with-sc-state}
	\end{figure}

  In Fig.\,\ref{fig:her-overlap-plot-with-sc-state}, deviation of the resulting fidelity  from one is plotted as a function of \(\beta\). It does not exceed $3 \cdot 10^{-3}$ and most probably results from the accumulated errors of the numerical calculations (the dip in the vicinity of \(\beta = 57\) and several smaller dips in at \(\beta = 30\text{-}50\), most probably, originate for the peculiarities of the numerical calculation library we used).

	If necessary, then the prepared squeezed SC state $\ket{\psi_{\rm out\,raw}}$ could be converted to the non-squeezed one $\ket{\psi_{\rm out}}$ using the squeezing operation with the negative squeeze factor \(-\xi_{\rm prep}\), see Eq.\,\eqref{psi_out}, implemented by means of a degenerate optical parametric oscillator in the  strong inexhaustible classical pumping mode regime, see {\it e.g.} the review \cite{Schnabel_PR_684_1_2017}. The  resulting average photon number can be calculated as follows (see Eq.\,\eqref{eq:energy-sc-neg-sq}):
	\begin{equation}
		N = 2 \beta^2 e^{2 \xi_{\rm prep}} - \tanh(2\alpha^2) \sinh^2(\xi_{\rm prep}) e^{2\xi_{\rm prep}}
	\end{equation}
	where \(\xi_{\rm prep}\) is given by Eq.\,\eqref{eq:sqeueeze-factor-approx}).

  Also if necessary, the tilt of the prepared state can be adjusted using the rotation (that is the phase delay) operation.

	\section{Discussions and conclusions}\label{sec:possible-implementation}

  We have shown that combining the degenerate parametric down-conversion process with the heralding measurement of the photon number in the pump mode, it is possible to prepare the signal mode in a bright quantum state that is virtually indistinguishable from the squeezed Schr\"odinger cat state. If necessary, it can be converted to the ``ordinary'' non-squeezed state by using the subsequent standard anti-squeezing operation.

	Concerning the practical implementation of this method in the optical band, the dielectric optical microresonators \cite{V_B_Braginsky_PRA_137_7_1989, D_V_Strekalov_JournalOfOptics_18_12_2016} that combine very low optical losses with a high concentration of the optical energy in the small volumes of their optical modes look as a promising platform.

  The evident necessary condition for our protocol is that the preparation time $t_{\rm opt}$, see Eq.\,\eqref{eq:t-max-by-beta} have to be shorter than optical mode relaxation time $t^* = Q/\omega_s$, where $Q$ is the quality factor of the mode.  The values of intrinsic quality factors of the modern lithium niobate microresonators could exceed $10^8$, see e.g. Refs.\,\cite{V_S_Ilchenko_PRL_92_4_2004,  G_Conti_OptExpress_19_4_2011, R_Gao_NewJournalOfPhys_23_12_2021}. At the wavelength $1.55\,\mu\text{m}$, this $Q$ correspond to the relaxation time of about
  \begin{equation}\label{t_star}
    t^* \simeq 10^{-7}\,{\rm s} \,.
  \end{equation}

  The factor $\gamma$, in the case of isotropic medium and rectangular spatial distribution of the optical fields, can be calculated as follow, see Refs.\,\cite{R_W_Boyd_Nonlinear_Optics_Book_2008, C_Okoth_PRA_99_4_2019}:
 		\begin{equation}\label{gamma}
 			\gamma =
 			\frac{\chi^{(2)} \omega_s}{4\sqrt{2} n_s^2 n_p}
 			\sqrt{\frac{\hbar \omega_p}{V \epsilon_0}} \,,
 		\end{equation}
 		where $\chi^{(2)}$ is the non-linearity factor, \(n_s, n_p\) are the refraction indices for, respectively, the signal and pump modes, $V$ is the volume of the modes, and $\epsilon_0$ is vacuum permittivity.

  Using the typical for the sub-mm resonators value of $V\sim10^{-15}\,{\rm m}$ and assuming the signal wavelength equal to $1.55\,\mu\text{m}$, we obtain the following estimate for $\gamma$:
 		\begin{equation}\label{gamma_num}
 			\gamma \simeq 2\times10^6\,\text{s}^{-1} \,,
 		\end{equation}
 		For the values of $\beta\gtrsim10$, this value translates to quantum state preparation time (see Eq.\,\eqref{eq:t-max-by-beta}) equal to
    \begin{equation}\label{t_opt_est}
      t_{\rm opt} \simeq \Bigl(\frac{10}{\beta}\Bigr)^{0.84}\times10^{-7}\,{\rm s} \,.
    \end{equation}

  It follows from of Eqs.\,\eqref{t_star} and \eqref{t_opt_est} that taking into account the rapid progress in technologies of fabrication of the optical microresonators, the method proposed in this work can be considered as a promising option for preparation of bright (with $\beta\gg1$) Schr\"odinger cat states.

	\acknowledgments

	This work was supported by the Russian Science Foundation (project 20-12-00344) and Theoretical Physics and Mathematics Advancement Foundation “BASIS” (Grant 23-1-1-39-1). The author would like to thank D.\,Chermoshentsev, B.\,Nougmanov, and F.\,Khalili for useful discussions.

	\appendix
	\subsection{Derivation of Eq.\,\eqref{eq:amplitude-of-prep-sc}}\label{sec:app-sc-state-ampl-deriv}
	Initial state before parametric interaction is \(\psi_0 = \ket{0}_{\rm s} \otimes \ket{\beta}_{\rm p}\). This state energy equals \(E_0 = 2 \beta^2\), where we assumed that \(\beta\) is real.

	According the proposed in this paper procedure output state is (we rename \(\xi_{\rm prep}\) as \(\xi\) for brevity):
	\begin{equation}
		\ket{\psi} = \frac{1}{\sqrt{K}} \hat{S}(\xi) \left(\ket{\alpha} + \ket{-\alpha}\right) \otimes \ket{0}_{\rm p}\,,
	\end{equation}
	where \(K = 2(1 + e^{-2\alpha^2})\) and we assume that \(\alpha\) is real.

	Energy of prepared state \(E\) could be found as:
	\begin{equation}
		E = \frac{1}{K} \left(\bra{\alpha} + \bra{-\alpha}\right) \hat{S}^\dagger (\xi)\hat{a}^\dagger \hat{a} \hat{S}(\xi) \left(\ket{\alpha} + \ket{-\alpha}\right)
	\end{equation}

	We will use well-known property of squeeze operator:
	{\allowdisplaybreaks
	\begin{gather*}
		\hat{S}^\dagger(\xi) \hat{a} \hat{S}(\xi) = \hat{a} \cosh(r) - e^{i \theta} \hat{a}^\dagger \sinh(r)
		\\
		\hat{S}^\dagger(\xi) \hat{a}^\dagger \hat{S}(\xi) = \hat{a}^\dagger \cosh(r) - e^{-i \theta} \hat{a} \sinh(r)\,,
	\end{gather*}
	}
	where \(\xi = r e^{i \theta},\, r,\theta \in \mathbb{R}\). As a result, we obtain:
	\begin{equation}
	\begin{aligned}
		\hat{S}^\dagger (\xi)\hat{a}^\dagger \hat{a} \hat{S}(\xi)
		=
		\hat{S}^\dagger (\xi)\hat{a}^\dagger \hat{S}(\xi) \hat{S}^\dagger \hat{a} \hat{S}(\xi)
		= \\ =
		\cosh(2r) \hat{a}^\dagger \hat{a} + \sinh^2(r)
		-
		\frac{1}{2} \sinh(2r) \left(e^{i \theta} \hat{a}^{\dagger 2} + e^{- i \theta} \hat{a}^2\right)
	\end{aligned}
	\end{equation}
	If \(\xi < 0\), then
	\begin{equation}
	\begin{aligned}
		\hat{S}^\dagger (\xi)\hat{a}^\dagger \hat{a} \hat{S}(\xi)
		=
		\cosh(2r) \hat{a}^\dagger \hat{a} + \sinh^2(r)
		+
		\frac{1}{2} \sinh(2r) \left(\hat{a}^{\dagger 2} +\hat{a}^2\right)
	\end{aligned}
	\end{equation}
	Therefore, the total energy is equal to:
	\begin{equation}
	\begin{aligned}\label{eq:energy-sc-neg-sq}
		E =
		\tanh(2 \alpha^2) \left(\alpha^2 e^{2r} + \sinh^2(r) \right)
	\end{aligned}
	\end{equation}
Using energy preservation: \(E_0 = E\), we obtain:
	\begin{equation}
		\alpha = \frac{ e^{-r}}{\tanh^2(2 \alpha^2)} \sqrt{2 \beta^2 - \sinh^2(r)}
	\end{equation}

\end{document}